# A Memetic Algorithm for the Multidimensional Assignment Problem

G. Gutin[1] and D. Karapetyan[1]

Royal Holloway, University of London,
gutin@cs.rhul.ac.uk,
daniel.karapetyan@gmail.com

**Abstract.** The Multidimensional Assignment Problem (MAP or $s$-AP in the case of $s$ dimensions) is an extension of the well-known assignment problem. The most studied case of MAP is 3-AP, though the problems with larger values of $s$ have also a number of applications. In this paper we propose a memetic algorithm for MAP that is a combination of a genetic algorithm with a local search procedure. The main contribution of the paper is an idea of dynamically adjusted generation size, that yields an outstanding flexibility of the algorithm to perform well for both small and large fixed running times. The results of computational experiments for several instance families show that the proposed algorithm produces solutions of very high quality in a reasonable time and outperforms the state-of-the art 3-AP memetic algorithm.

## 1 Introduction

The Multidimensional Assignment Problem (MAP or $s$-AP in the case of $s$ dimensions) is an extension of the well-known Assignment Problem (AP, linear AP) which is exactly two dimensional case of MAP. MAP has a host of applications, see, e.g., [1, 3, 4] for 'classic' applications and [2, 12, 13] for recent applications in solving systems of polynomial equations and centralized multisensor multitarget tracking. In fact, several applications described in [2, 3, 12, 13] naturally require the use of $s$-AP for values of $s$ larger than 3.

For a fixed $s \geq 2$, $s$-AP is stated as follows. Let $X_1 = X_2 = \ldots = X_s = \{1, 2, \ldots, n\}$ and let $X = X_1 \times X_2 \times \ldots \times X_s$. For a vector $e \in X$, the component $e_j$ denotes its $j$th coordinate, i.e., $e_j \in X_j$. Each vector $e \in X$ is assigned a non-negative weight $w(e)$. A collection $A$ of $t \leq n$ vectors $e^1, e^2, \ldots, e^t$ is a *(feasible) partial assignment* if $e_j^i \neq e_j^k$ holds for each $i \neq k$ and $j \in \{1, 2, \ldots, s\}$. The *weight* of a partial assignment $A$ is $w(A) = \sum_{i=1}^{t} w(e^i)$. An *assignment* (or *full assignment*) is a partial assignment with $n$ vectors. The objective of $s$-AP is to find an assignment of minimum weight.

While AP can be solved in a polynomial time [10], $s$-AP for every $s \geq 3$ is NP-hard [5]. MAP is a very hard problem in the following sense. The weight matrix of MAP contains $n^s$ values, there exist $n!^{s-1}$ possible assignments and the fastest known algorithm to find the optimal one takes $O(n!^{s-2}n^3)$ time. Indeed, without loss of generality set $e_1^i = i$ for $i = 1, 2, \ldots, n$ and find the optimal values of $e_s^i$ by solving a linear AP for every possible combination of the values $e_j^i$ for $i = 1, 2, \ldots, n$ and $j = 2, 3, \ldots, s-1$ such that $e_j^i \neq e_j^k$ for $i \neq k$. Compare it with, e.g., the Travelling Salesman Problem which has only $n^2$ weights, $(n-1)!$ possible tours and that can be solved in $O(n^2 2^n)$ time.

The rest of the paper is organized as follows. The general scheme and all the details of the proposed memetic algorithm including the local search procedure are described in Section 2. The test bed for our computational evaluation is discussed in Section 3. The experiments with our and some other metaheuristics are described and the results are discussed in Section 4. The main outcomes of the presented research are reported in Section 5.

## 2 The Algorithm

A memetic algorithm is a combination of genetic algorithm with local search. A typical scheme of a memetic algorithm is as follows.

1. Produce the first generation, i.e., a set of solutions.
2. Apply the local search procedure to every solution in the first generation.
3. Repeat the following while a termination criterion is not met:
   (a) Produce a set of new solutions by applying so-called genetic operators to solutions from the previous generation.
   (b) Improve every solution in this set with the local search procedure.
   (c) Select several best solutions from this set to the next generation.

While the general scheme of the algorithm is quite common for all memetic algorithms, the set of genetic operators and the way they are applied can vary significantly. In our algorithm, we use the following procedure to obtain the next generation:

$$g^{i+1} = selection(\{g_1^i\} \cup mutation(g^i \setminus \{g_1^i\}, p_m, \mu_m) \cup C),$$

where $g^k$ is the $k$th generation and $g_j^k$ is the $j$th assignment of the $k$th generation; $g_1^k$ is the best assignment in the $k$th generation. Constants $p_m = 0.5$ and $\mu_m = 0.1$ define the probability and the strength of mutation operator respectively. The function *selection* simply returns $m_{i+1}$ best distinct assignments among the given ones, where $m_k$ is the size of the $k$th generation (if the number of distinct assignments in the given set is less than $m_{i+1}$, *selection* returns all the distinct assignments and updates the value of $m_{i+1}$ accordingly). The set of assignments $C$ (crossover part) is calculated as follows:

$$C = \bigcup_{j=1}^{(p \cdot m_{i+1} - m_i)/2} LocalSearch(crossover(g_u^i, g_v^i))$$

where $u, v \in \{1, 2, \ldots, m_i\}$ are chosen randomly for every *crossover* run and $p = 3$ defines how many times more assignments should be produced for the selection operator. The mutation function for a set of solutions is defined as follows:

$$mutation(G, p, \mu) = \bigcup_{g \in G} \begin{cases} LocalSearch(perturb(g, \mu)) & \text{if } r < p \\ g & \text{otherwise} \end{cases}$$

where $r \in [0, 1]$ is chosen randomly every time. The functions $crossover(x, y)$, $perturb(x, \mu)$ and $LocalSearch(x)$ are discussed later.

### 2.1 Coding

In genetic algorithms, coding is a way to represent a solution as a sequence of atom values such as boolean values or numbers; genetic operators are applied to such sequences. Good coding should meet the following requirements:

- Coding should be a bijection, i.e., a decoding procedure such as $decode(code(x)) = x$ for any feasible solution $x$ should exist.
- Evaluation of the quality of a coded solution should be fast.
- Every fragment of the coded solution should refer to just a part of the whole solution, so a small change in the coded sequence should not change the whole solution.
- It should be relatively easy to produce algorithms for random modification of a solution (mutation) and for combination of two solutions (crossover) which produce feasible solutions.

Huang and Lim [7] use a local search procedure that, having first two dimensions of an assignment, determines the third dimension (note that the algorithm from [7] is designed only for 3-AP). Since the first dimension can always be fixed with no loss of generality, one needs to store only the second dimension of an assignment. Unfortunately, this coding requires a specific local search and is robust for 3-AP only. We use a different coding: a vector of an assignment is considered as an atom in our algorithm and, thus, a coded assignment is just a list of its vectors. The vectors are always stored in the first coordinate ascending order, e.g., an assignment

$$\{(2, 4, 1), (4, 3, 4), (3, 1, 3), (1, 2, 2)\}$$

would be represented as

$$(1, 2, 2), (2, 4, 1), (3, 1, 3), (4, 3, 4).$$

### 2.2 Termination Condition

Usually, a termination condition in a memetic algorithm tries to predict the point after which any further effort is useless or, at least, not efficient. A typical approach is to count the number of subsequent generations which did not improve the best result and to stop the algorithm when this number reaches some predefined value.

We use a different approach. To be able to compare different algorithms correctly and to satisfy real world requirements, we bound our algorithm within some fixed running time. Apart from the mentioned advantages of this termination condition, it is worth to note that it gives flexibility to produce either fast or high quality solutions depending on one's needs.

### 2.3 Generation Size

The most natural way to fit the running time of a memetic algorithm into the given bound is to produce generations of some fixed size until the time is elapsed. However, it is clear that one memetic algorithm cannot work efficiently in both cases if there are just a few generations and if there are hundreds of generations. Thus, instead, we fix the number of generations and vary the generation size.

Our computational experiments show that, with a fixed running time, the most appropriate number $I$ of generations for our algorithm is always around 50; this number does not depend on the local search procedure or the given time. Since the running time of the local search procedure can vary significantly (e.g., last generations usually contain better solutions than the first ones and, thus, are processed faster) and also to make our algorithm easily portable, we decided to adjust the generation size dynamically according to the remained time such that the total number of generations would always be close to $I$.

In particular, the size of the next generation is calculated as follows:

$$m'_{i+1} = \begin{cases} m'_i \cdot \max\left\{\min\left\{\frac{T-t}{\Delta \cdot (I-i)}, k\right\}, \frac{1}{k}\right\} & \text{if } i < I \\ m'_i \cdot k & \text{otherwise} \end{cases},$$

where $T$ is the given time, $t$ is the elapsed time, $\Delta$ is the time spent to produce the previous generation, $I$ is the prescribed number of generations and $k = 1.25$ is a constant that limits the generation size change. The 'otherwise' branch normally should not be used but in case of quick finish of the $I$th generation it increases the size of generation (recall that the goal is to keep the number of generations close to $I$). Note that the values $m'_i$ are real numbers, and the actual size $m_i$ of the $i$th generation is defined as a rounded down value of $m'_i$. In addition, the value $p \cdot m_{i+1} - m_i$, i.e., the number of assignments produced by crossover, should always be even because every crossover produces exactly two assignments. If this condition is not met, the value of $m_{i+1}$ is increased by one:

$$m_i = \max\left\{4, \begin{cases} \lfloor m'_i \rfloor & \text{if } (p \cdot \lfloor m'_i \rfloor - m_{i-1}) \text{ is even} \\ \lfloor m'_i \rfloor + 1 & \text{otherwise} \end{cases}\right\},$$

The size of the first generation is obtained in a different way (see Section 2.4).

### 2.4 First Generation

As it was shown in [6] (and we also confirmed it by experimentation with our memetic algorithm and the construction heuristics from [8]), it is good to start any MAP local search or metaheuristic from a Greedy construction heuristic. Thus, we start from running the Greedy algorithm (we use the same implementation as in [6]) and then perturb it using our *perturb* procedure to obtain every item of the first generation:

$$g_j^1 = LocalSearch(perturb(greedy, \mu_f)),$$

where *greedy* is an assignment obtained by the Greedy heuristic and $\mu_f = 0.2$ is the perturbation strength coefficient. Since *perturb* performs a random modification, it guarantees some diversity in the first generation.

The algorithm produces assignments for the first generation until $T/I$ time elapses or at least 4 assignments (recall that $T$ is the time given for the whole memetic algorithm and $I$ is the prescribed number of generations).

### 2.5 Crossover

A typical crossover operator combines two solutions, parents, to produce two new solutions, children. Crossover is the main genetic operator, i.e., it is the source of a genetic algorithm power. Due to the selection operator, it is assumed that good fragments of solutions are spread wider than others and that is why, if both parents have some similar fragments, these fragments are probably good and should be copied without any change to the children solutions. Other parts of the solution can be randomly mixed and modified though they should not be totally destroyed.

The widest used crossover is the one-point crossover that produces two children $x'$ and $y'$ from two parents $x$ and $y$ as follows: $x'_i = x_i$ and $y'_i = y_i$ for every $i = 1, 2, \ldots, k$ and $x'_i = y_i$ and $y'_i = x_i$ for every $i = k+1, k+2, \ldots, n$, where $k \in \{1, 2, \ldots, n-1\}$ is chosen randomly. One can see that if $x_i = y_i$ for some $i$, then the corresponding values in the children sequences will be the same: $x'_i = y'_i = x_i = y_i$.

However, the one-point and some other standard crossovers do not preserve feasibility of MAP assignments since not every sequence of vectors can be decoded into a feasible assignment. We propose a special crossover operator. Let $x$ and $y$ be the parent assignments and $x'$ and $y'$ be the child assignments. First, we retrieve equal vectors in the parent assignments and initialize both children with this set of vectors:

$$x' = y' = x \cap y .$$

Let $k = |x \cap y|$, i.e., the number of equal vectors in the parent assignments, $p = x \setminus x'$ and $r = y \setminus y'$, where $p$ and $r$ are ordered sets. Let $\pi$ and $\omega$ be random permutations of size $n - k$. For every $j = 1, 2, \ldots, n - k$ the crossover sets either $x' = x' \cup p_{\pi(j)}$ and $y' = y' \cup r_{\omega(j)}$ or $x' = x' \cup r_{\omega(j)}$ and $y' = y' \cup p_{\pi(j)}$. The particular option is chosen randomly with the probability of the first one 80%.

Since this procedure can yield infeasible assignments, it requires additional correction of the child solutions. For this purpose, the following is performed for every dimension $d = 1, 2, \ldots, s$ for every child assignment $c$. For every $i$ such that $\exists j < i : c_d^j = c_d^i$ set $c_d^i = r$ where $r \in \{1, 2, \ldots, n\} \setminus \{c_d^1, c_d^2, \ldots, c_d^n\}$ is chosen randomly. In the end of the correction procedure, the vectors in the assignment should be sorted in the ascending order of the first coordinates (see Section 2.1).

In other words, our crossover copies all equal vectors from the parent assignments to the child ones, then copies the rest of the vectors randomly choosing every time a pair of vectors, one from the first parent and one from the second one, and then adding them either to the first and to the second child respectively or, with probability 20%, to the second and to the first child respectively. Since the obtained child assignments can be infeasible, the crossover corrects each of them; for every dimension of every child it replaces all duplicate coordinates with randomly chosen correct ones, i.e., with the coordinates which are not currently used for that dimension.

### 2.6 Perturbation Algorithm

The perturbation procedure $perturb(x, \mu)$ is intended to modify randomly an assignment $x$ with the given strength $\mu$. In our memetic algorithm, perturbation is used to produce the first generation and to mutate assignments from the previous generation when producing the next generation.

Our perturbation procedure $perturb(x, \mu)$ performs $\lceil n\mu/2 \rceil$ random swaps. Each swap selects two vectors and some dimension randomly and then swaps the corresponding coordinates: swap $x_u^d$ and $y_v^d$, where $u, v \in \{1, 2, \ldots, n\}$ and $d \in \{1, 2, \ldots, s\}$ are chosen randomly; repeat the procedure $\lceil n\mu/2 \rceil$ times. For example, if $\mu = 1$, our perturbation procedure modifies up to $n$ vectors in the given assignment.

### 2.7 Local Search Procedure

An extensive study of a number of local search heuristics for MAP can be found in [6]; the paper includes both fast and slow algorithms. It also shows that a combination of two heuristics can yield a heuristic superior to the original ones.

The following heuristics were considered as possible candidates for the local search procedure for our memetic algorithm (we provide only a brief description of every heuristic here; full descriptions can be found in [6]):

- 2-opt is a simple heuristic that selects the best of all possible interchanges for every pair of vectors in the assignment. 2-opt is known as a very fast but poor quality heuristic.
- DV is a natural extension of a local search proposed and used in [7]. On every iteration, DV fixes all but one dimensions yielding a 2-AP instance, solves it (recall that 2-AP can be solved in a polynomial time) and reorders the unfixed dimension according to the obtained solution of 2-AP.

- MDV is a modification of DV proposed in [6]. While DV divides the dimension set into $\{1, 2, \ldots, i-1, i+1, \ldots s\}$ and $\{i\}$ for every $i$, MDV tries every possible division, fixes dimensions from every of two sets together yielding a 2-AP and then proceeds as DV. For 3-AP, DV and MDV are equivalent while for $s$-AP with $s > 3$ MDV is more powerful.
- DV2 is a combination of 2-opt and DV; it applies sequentially 2-opt and DV to the given assignment. DV2 repeats the whole procedure until one of the heuristics appears to be incapable to improve the assignment.
- MDV2 is like DV2 but instead of DV it applies MDV.
- MDV3 is like MDV2 but instead of 2-opt it applies 3-opt. 3-opt, similarly to 2-opt, tries every possible interchange for every triple of vectors in the assignment. While 3-opt is a very slow and not very successful heuristic, MDV3 is relatively fast and outperforms all other considered local search heuristics with respect to solution quality.
- MDVV is like MDV2 but instead of 2-opt is applies v-opt. v-opt is an extension for $s$-AP with arbitrary $s$ of the Variable Depth Interchange heuristic proposed in [1] for 3-AP. Like 2-opt, v-opt considers interchanges of vector pairs, however the objective and the enumeration order in v-opt are totally different.

Results for 3-opt and v-opt as a local search for our memetic algorithm are not given in this paper since they did not show any promising results in our experiments.

Table 1 shows comparison of the results of our memetic algorithm based on different local search procedures. One can see that the fast heuristics MDV, DV2 and MDV2 [6] perform better than others in almost every experiment, and MDV2 shows the best average result among them. Similar results were obtained for the Multichain metaheuristic proposed in [6] (the results reported in [6] are slightly different but it can be explained by the difference in the test beds). The situation stays approximately the same in the experiments with different given times. Thus, MDV2 was selected as a local search procedure for our memetic algorithm.

## 3 Test Bed

In this section we discuss instance families used for experimental evaluation of our memetic algorithm. We use pseudo real-world instances: Composite Cyclic (CC), Composite Clique (CQ), Squares Root (SR) and perturbed versions of each of them. The CC and CQ families were proposed before in the literature (see [3] and references there) while the others are new.

To explain these instance families, it is better to represent MAP as a graph problem. Let $G = (V_1 \cup V_2 \cup \ldots \cup V_s, E)$ be an $s$-partite graph where $|V_i| = n$ for every $i = 1, 2, \ldots, s$. A weight is assigned to every clique in $G$, i.e., to every induced subgraph $(\{v_1, v_2, \ldots, v_s\}, \{v_1v_2, v_1v_3, \ldots, v_1v_s, v_2v_3, v_2v_4, \ldots, v_2v_s, \ldots, v_{s-1}v_s\})$, where $v_1 \in V_1$, $v_2 \in V_2$, $\ldots$, $v_s \in V_s$; a clique corresponds to a vector in the original interpretation. The objective is to find $n$ disjoint cliques such that their total weight is minimized.

The idea of all CC, CQ and SR is that in the real life it is usually possible to define some relation only between two objects but not between $s$ objects. Thus, the graph $G$ should be weighted and the weight of a clique $C = (V_C, E_C)$ should be a function of the weights $E_C$ of the edges in the clique. We immediately get two natural instance families, CC and CQ:

$$w_{CC}(E_C) = w(v_s v_1) + \sum_{i=1}^{s-1} w(v_i v_{i+1}) \ ,$$

$$w_{CQ}(E_C) = \sum_{e \in E_C} w(e) \ .$$

The SR instance family, as well as CC, is based on the weights in a cycle:

$$w_{CC}(E_C) = \sqrt{w^2(v_s v_1) + \sum_{i=1}^{s-1} w^2(v_i v_{i+1})} \ .$$

In this case, the objective is not only to minimize the considered weights but also to keep all the weights not too large.

The perturbed instances are the same instances but the weight assigned to every clique is modified randomly: $w(c) = w(c) + r$ for every clique $c$, where $r \in \{0, 1, \ldots, 19\}$ is chosen randomly.

Our test bed includes instances of 3-AP, 4-AP, 5-AP and 6-AP, six types of instances per each value of $s$: CC, perturbed CC, CQ, perturbed CQ, SR and perturbed SR (for $s = 3$, CQ and perturbed CQ were excluded from the test bed since CC and CQ for 3-AP are equivalent; however, the results for the 3-AP CC (perturbed CC) instances are included in the CQ (perturbed CQ) averages). The sizes $n$ of the instances are as follows: $n = 40$ for 3-AP, $n = 30$ for 4-AP, $n = 18$ for 5-AP and $n = 12$ for 6-AP. These sizes were obtained empirically such that the difficulty of all instances in the test bed is approximately the same (in other words, a local search takes approximately equal times for every considered instance). We produced 10 different instances for every experiment, i.e., for every combination of $s$, $n$ and instance family; the results reported in the tables in Section 4 in every row are averages among 10 runs. To generate the instances, we initialized every edge of the graph $G$ with a random number from $\{1, 2, \ldots, 100\}$ using standard Miscrosoft .NET random generator [11] which is based on the Donald E. Knuth's subtractive random number generator algorithm [9]. As a seed of the random number sequences for all the instance types we use the following number: $seed = s + n + i$, where $i \in \{1, 2, \ldots, 10\}$ is an index of the instance of this type and size.

## 4 Experiment Results

Three metaheuristics were considered in our experiments:

- A Multichain metaheuristic (MC) proposed in [6] which significantly outperforms all local search heuristics known from the literature with respect to solution quality.
- An extended version of the memetic algorithm by Huang and Lim [7] (HL).
- Our memetic algorithm (GK).

For the purpose of comparison, the Huang and Lim's heuristic was extended for $s > 3$ as follows:

- The coded assignment contains not only the second dimension but it stores sequentially all the dimensions except the first and the last ones, i.e., an assignment $\{e^1, e^2, \ldots, e^s\}$ is represented as $e_2^1, e_2^2, \ldots, e_2^n, e_3^1, e_3^2, \ldots, e_3^n$, $\ldots$, $e_{s-1}^1, e_{s-1}^2, \ldots, e_{s-1}^n$ ($e_1^i = 1$ for each $i$ and $e_s^i$ can be chosen in an optimal way solving an AP).
- The local search heuristic, that was initially designed for 3-AP, is extended to DV as it is described in [6].
- The crossover, proposed in [7], is applied separately to every dimension (except the first and the last ones) since it is designed for one dimension only (recall that the memetic algorithm from [7] stores only the second dimension of an assignment, see Section 2.1).

Our computational experience show that the solution quality of our implementation of the Huang and Lim's heuristic is similar to the results reported in [7] and the running time is slightly larger because of the extension for $s > 3$.

All the heuristics are implemented in Visual C++ and evaluated on a platform based on an AMD Athlon 64 X2 3.0 GHz processor.

The results are reported in Tables 2 and 3. The following running times are considered in this tables: 0.3 s, 1 s, 3 s, 10 s and 30 s. The first given time, 0.3 s, corresponds approximately to the running time of HL if it applies its original termination condition. Every entry of these tables contains an average solution error for 10 instances of some fixed type and size but of different seed values (see Section 3 for details). The value of the solution error is calculated as $(v - v_{\text{best}})/v_{\text{best}} \cdot 100\%$, where $v$ is the obtained solution value and $v_{\text{best}}$ is the best known solution value[1] (the best solution values can be found in Table 1). The instance name consists of three parts: the number of dimensions $s$, the type of the instance and the size $n$ of the instance. In addition, letter 'p' is added to the instance name to indicate that it is a perturbed instance.

The average values for different instance families and numbers of dimensions are provided at the bottom of each table. The CQ (perturbed CQ) averages include CC (perturbed CC) results for 3-AP since CC and CQ instance families are equivalent when $s = 3$ and, thus, CQ and perturbed CQ instances for 3-AP are excluded from the test bed. The best among MC, HL and GK results are underlined in every row for every particular given time.

One can see that GK clearly outperforms MC and HL: GK performs better than others in most experiments both with small given times (0.3 and 1 seconds) and, especially, with larger given times (3, 10 and 30 seconds) for which it is the best in every experiment with just a few exceptions. GK shows the best average results for every instance family, for every number of dimensions $s$ and for every given time. The latter fact proves its ability to perform well for very different given times. The MC heuristic mostly outperforms HL for small given times but shows approximately the same or slightly worse results for larger given times that was expected because memetic algorithms are normally intended to be high quality but slow heuristics.

---

[1] The best known solutions were obtained during our experiments with different heuristics.

It is worth to note that we experimented with different values of the GK algorithm parameters such as $I$, $\mu_f$, $\mu_m$ etc. and concluded that small variations of these values do not significantly influence the algorithm performance.

## 5 Conclusion

A memetic algorithm for MAP is proposed in this paper. The main feature of the algorithm is a dynamically adjusted generation size. It was shown that there exists the most appropriate number of generations which does not depend on the given instance or solution quality requirements. This idea allowed us to design a flexible algorithm that is able to perform well for very different given times. Extensive experiments show that the proposed algorithm clearly outperforms all high quality MAP algorithms known from the literature given the same running time.

**Table 1.** Comparison of the memetic algorithm based on different local search. The given time is 1 sec.

|  |  | Solution error, % | | | | | | |
|---|---|---|---|---|---|---|---|---|
| Inst. | Best | 2-opt | DV | MDV | DV2 | MDV2 | MDV3 | MDVV |
| 3cc40 | 926.9 | 9.04 | 1.12 | 0.73 | 0.58 | <u>0.54</u> | 4.51 | 0.93 |
| 3cc40p | 1221.4 | 5.85 | 0.66 | 0.96 | 0.56 | <u>0.48</u> | 3.86 | 0.90 |
| 3sr40 | 610.6 | 9.11 | 0.79 | 1.31 | <u>0.41</u> | 0.47 | 5.00 | 1.15 |
| 3sr40p | 862.7 | 8.59 | 0.51 | 0.80 | <u>0.35</u> | 1.09 | 5.12 | 0.95 |
| 4cc30 | 921.0 | 9.75 | 5.56 | 2.89 | 2.48 | <u>1.41</u> | 9.09 | 5.82 |
| 4cc30p | 1126.3 | 9.85 | 7.04 | 3.78 | 2.90 | <u>1.06</u> | 9.39 | 7.34 |
| 4cq30 | 2281.9 | 4.15 | 2.06 | 1.96 | <u>1.17</u> | 1.41 | 6.67 | 3.80 |
| 4cq30p | 2529.2 | 5.52 | 1.88 | 1.59 | 1.12 | <u>0.91</u> | 7.67 | 3.32 |
| 4sr30 | 535.6 | 16.92 | 9.50 | 4.26 | 3.12 | <u>2.04</u> | 11.58 | 6.85 |
| 4sr30p | 706.5 | 13.64 | 8.58 | 4.61 | 5.51 | <u>2.35</u> | 11.22 | 7.53 |
| 5cc18 | 956.8 | 1.43 | 2.33 | 0.77 | 0.34 | <u>0.29</u> | 7.13 | 1.40 |
| 5cc18p | 1083.3 | 2.66 | 2.44 | 0.76 | 1.04 | <u>0.18</u> | 7.38 | 1.98 |
| 5cq18 | 3458.6 | 0.94 | 0.17 | 0.15 | 0.10 | <u>0.05</u> | 3.30 | 1.72 |
| 5cq18p | 3617.0 | 1.16 | 0.24 | 0.60 | <u>0.14</u> | 0.24 | 3.58 | 0.94 |
| 5sr18 | 504.9 | 4.08 | 3.37 | 0.85 | 0.87 | <u>0.34</u> | 8.69 | 1.86 |
| 5sr18p | 610.5 | 4.73 | 4.46 | 2.62 | 2.24 | <u>0.84</u> | 12.33 | 3.03 |
| 6cc12 | 1023.8 | 0.24 | 0.79 | <u>0.00</u> | 0.17 | 0.12 | 3.87 | 0.43 |
| 6cc12p | 1108.3 | 0.17 | 0.59 | 0.10 | 0.13 | <u>0.03</u> | 3.46 | 0.22 |
| 6cq12 | 4505.6 | 0.28 | 0.16 | 0.26 | <u>0.00</u> | 0.30 | 3.51 | 1.17 |
| 6cq12p | 4608.9 | 0.16 | 0.07 | 0.31 | <u>0.00</u> | 0.22 | 2.69 | 0.51 |
| 6sr12 | 502.9 | 0.50 | 0.68 | <u>0.02</u> | 0.08 | 0.04 | 9.33 | 0.46 |
| 6sr12p | 567.8 | 0.46 | 1.02 | 0.35 | 0.18 | <u>0.05</u> | 5.30 | 1.95 |
| All avg. | | 4.96 | 2.45 | 1.35 | 1.07 | <u>0.66</u> | 6.58 | 2.47 |
| CC avg. | | 5.12 | 2.45 | 1.10 | 0.89 | <u>0.59</u> | 6.15 | 2.14 |
| CC p. avg. | | 4.63 | 2.68 | 1.40 | 1.16 | <u>0.44</u> | 6.02 | 2.61 |
| CQ avg. | | 3.60 | 0.88 | 0.78 | <u>0.46</u> | 0.57 | 4.50 | 1.90 |
| CQ p. avg. | | 3.17 | 0.71 | 0.86 | <u>0.45</u> | 0.46 | 4.45 | 1.42 |
| SR avg. | | 7.65 | 3.58 | 1.61 | 1.12 | <u>0.72</u> | 8.65 | 2.58 |
| SR p. avg. | | 6.86 | 3.64 | 2.10 | 2.07 | <u>1.08</u> | 8.50 | 3.37 |
| 3-AP avg. | | 8.15 | 0.77 | 0.95 | <u>0.47</u> | 0.65 | 4.62 | 0.98 |
| 4-AP avg. | | 9.97 | 5.77 | 3.18 | 2.71 | <u>1.53</u> | 9.27 | 5.78 |
| 5-AP avg. | | 2.50 | 2.17 | 0.96 | 0.79 | <u>0.32</u> | 7.07 | 1.82 |
| 6-AP avg. | | 0.30 | 0.55 | 0.17 | <u>0.09</u> | 0.13 | 4.69 | 0.79 |

**Table 2.** Metaheuristics comparison.

|  | Solution error, % | | | | | | | | |
|---|---|---|---|---|---|---|---|---|---|
|  | 0.3 sec. | | | 1 sec. | | | 3 sec. | | |
| Inst. | HL | MC | GK | HL | MC | GK | HL | MC | GK |
| 3cc40 | 2.48 | 2.18 | <u>1.54</u> | 0.91 | 1.62 | <u>0.54</u> | 0.50 | 1.38 | <u>0.35</u> |
| 3cc40p | 2.25 | 1.32 | <u>1.15</u> | <u>0.48</u> | 0.97 | <u>0.48</u> | 0.44 | 0.79 | <u>0.34</u> |
| 3sr40 | 2.88 | 2.52 | <u>2.03</u> | 0.97 | 1.87 | <u>0.47</u> | 0.43 | 1.77 | <u>0.29</u> |
| 3sr40p | 3.62 | 2.84 | <u>2.19</u> | <u>0.88</u> | 2.02 | 1.09 | 0.58 | 1.56 | <u>0.23</u> |
| 4cc30 | 13.38 | <u>2.24</u> | 3.06 | 3.82 | <u>1.30</u> | 1.41 | 1.92 | 0.86 | <u>0.50</u> |
| 4cc30p | 13.16 | 4.43 | <u>4.35</u> | 5.47 | 3.50 | <u>1.06</u> | 2.03 | 2.60 | <u>0.62</u> |
| 4cq30 | 6.39 | 3.02 | <u>2.29</u> | 1.44 | 2.58 | <u>1.41</u> | 0.90 | 2.30 | <u>0.54</u> |
| 4cq30p | 5.91 | 2.73 | <u>2.33</u> | 1.58 | 1.90 | <u>0.91</u> | 1.10 | 1.36 | <u>0.73</u> |
| 4sr30 | 16.69 | 4.72 | <u>4.18</u> | 8.25 | 3.06 | <u>2.04</u> | 3.29 | <u>1.74</u> | 2.20 |
| 4sr30p | 18.95 | 6.57 | <u>4.71</u> | 7.71 | 4.16 | <u>2.35</u> | 2.94 | 3.18 | <u>0.93</u> |
| 5cc18 | 6.34 | 1.41 | <u>0.80</u> | 1.72 | 0.89 | <u>0.29</u> | 1.40 | 0.65 | <u>0.06</u> |
| 5cc18p | 7.13 | <u>2.01</u> | 2.05 | 2.22 | 1.60 | <u>0.18</u> | 1.96 | 1.60 | <u>0.28</u> |
| 5cq18 | 0.68 | 1.68 | <u>0.35</u> | 0.25 | 1.57 | <u>0.05</u> | 0.10 | 1.57 | <u>0.02</u> |
| 5cq18p | <u>0.61</u> | 1.50 | 0.63 | 0.29 | 1.14 | <u>0.24</u> | 0.23 | 0.67 | <u>0.09</u> |
| 5sr18 | 8.66 | 1.56 | <u>1.19</u> | 2.97 | 0.81 | <u>0.34</u> | 2.40 | 0.81 | <u>0.36</u> |
| 5sr18p | 10.19 | 3.72 | <u>2.69</u> | 3.62 | 3.60 | <u>0.84</u> | 2.47 | 3.60 | <u>0.29</u> |
| 6cc12 | 1.05 | 1.16 | <u>0.34</u> | 0.75 | 1.05 | <u>0.12</u> | 0.70 | 0.49 | <u>0.00</u> |
| 6cc12p | 2.17 | <u>0.22</u> | 0.29 | 1.10 | 0.22 | <u>0.03</u> | 0.76 | 0.22 | <u>0.00</u> |
| 6cq12 | <u>0.21</u> | 1.21 | 0.83 | <u>0.21</u> | 0.99 | 0.30 | 0.20 | 0.93 | <u>0.03</u> |
| 6cq12p | <u>0.10</u> | 0.89 | 0.28 | <u>0.10</u> | 0.86 | 0.22 | 0.07 | 0.86 | <u>0.02</u> |
| 6sr12 | 3.50 | 0.52 | <u>0.36</u> | 1.53 | 0.22 | <u>0.04</u> | 1.31 | 0.18 | <u>0.00</u> |
| 6sr12p | 5.64 | 1.44 | <u>1.02</u> | 1.90 | 0.74 | <u>0.05</u> | 1.90 | 0.67 | <u>0.00</u> |
| All avg. | 6.00 | 2.27 | <u>1.76</u> | 2.19 | 1.67 | <u>0.66</u> | 1.26 | 1.35 | <u>0.36</u> |
| CC avg. | 5.81 | 1.75 | <u>1.44</u> | 1.80 | 1.21 | <u>0.59</u> | 1.13 | 0.84 | <u>0.23</u> |
| CC p. avg. | 6.18 | 1.99 | <u>1.96</u> | 2.32 | 1.57 | <u>0.44</u> | 1.30 | 1.30 | <u>0.31</u> |
| CQ avg. | 2.44 | 2.02 | <u>1.25</u> | 0.70 | 1.69 | <u>0.57</u> | 0.42 | 1.54 | <u>0.23</u> |
| CQ p. avg. | 2.22 | 1.61 | <u>1.10</u> | 0.61 | 1.22 | <u>0.46</u> | 0.46 | 0.92 | <u>0.30</u> |
| SR avg. | 7.93 | 2.33 | <u>1.94</u> | 3.43 | 1.49 | <u>0.72</u> | 1.86 | 1.12 | <u>0.71</u> |
| SR p. avg. | 9.60 | 3.64 | <u>2.65</u> | 3.53 | 2.63 | <u>1.08</u> | 1.97 | 2.26 | <u>0.37</u> |
| 3-AP avg. | 2.81 | 2.21 | <u>1.73</u> | 0.81 | 1.62 | <u>0.65</u> | 0.49 | 1.38 | <u>0.30</u> |
| 4-AP avg. | 12.41 | 3.95 | <u>3.49</u> | 4.71 | 2.75 | <u>1.53</u> | 2.03 | 2.01 | <u>0.92</u> |
| 5-AP avg. | 5.60 | 1.98 | <u>1.28</u> | 1.85 | 1.60 | <u>0.32</u> | 1.43 | 1.48 | <u>0.18</u> |
| 6-AP avg. | 2.11 | 0.91 | <u>0.52</u> | 0.93 | 0.68 | <u>0.13</u> | 0.82 | 0.56 | <u>0.01</u> |

**Table 3.** Metaheuristics comparison.

|        | Solution error, % | | | | | |
|--------|---|---|---|---|---|---|
|        | 10 sec. | | | 30 sec. | | |
| Inst.  | HL | MC | GK | HL | MC | GK |
| 3cc40  | 0.42 | 1.05 | <u>0.22</u> | 0.39 | 0.73 | <u>0.02</u> |
| 3cc40p | 0.25 | 0.77 | <u>0.07</u> | 0.24 | 0.77 | <u>0.09</u> |
| 3sr40  | 0.28 | 1.42 | <u>0.03</u> | 0.28 | 1.08 | <u>0.00</u> |
| 3sr40p | 0.41 | 1.36 | <u>0.10</u> | 0.31 | 1.21 | <u>0.02</u> |
| 4cc30  | 1.55 | 0.53 | <u>0.23</u> | 1.43 | 0.39 | <u>0.10</u> |
| 4cc30p | 1.57 | 1.73 | <u>0.47</u> | 1.43 | 1.62 | <u>0.19</u> |
| 4cq30  | 0.85 | 1.70 | <u>0.52</u> | 0.74 | 1.47 | <u>0.11</u> |
| 4cq30p | 0.88 | 1.05 | <u>0.09</u> | 0.88 | 0.89 | <u>0.06</u> |
| 4sr30  | 3.04 | <u>0.84</u> | 0.99 | 2.74 | 0.58 | <u>0.11</u> |
| 4sr30p | 2.53 | 1.63 | <u>0.21</u> | 1.73 | 1.39 | <u>0.37</u> |
| 5cc18  | 1.00 | 0.65 | <u>0.00</u> | 0.85 | 0.65 | <u>0.00</u> |
| 5cc18p | 1.86 | 1.60 | <u>0.00</u> | 1.66 | 1.27 | <u>0.00</u> |
| 5cq18  | <u>0.04</u> | 1.57 | <u>0.04</u> | 0.04 | 1.57 | <u>0.02</u> |
| 5cq18p | 0.18 | 0.67 | <u>0.07</u> | 0.17 | 0.67 | <u>0.00</u> |
| 5sr18  | 1.96 | 0.81 | <u>0.16</u> | 1.96 | 0.81 | <u>0.00</u> |
| 5sr18p | 2.26 | 3.60 | <u>0.13</u> | 1.75 | 3.36 | <u>0.00</u> |
| 6cc12  | 0.70 | 0.40 | <u>0.00</u> | 0.70 | 0.32 | <u>0.00</u> |
| 6cc12p | 0.62 | 0.22 | <u>0.00</u> | 0.62 | 0.22 | <u>0.00</u> |
| 6cq12  | 0.20 | 0.93 | <u>0.00</u> | 0.20 | 0.93 | <u>0.00</u> |
| 6cq12p | 0.04 | 0.86 | <u>0.00</u> | 0.04 | 0.86 | <u>0.00</u> |
| 6sr12  | 1.11 | 0.04 | <u>0.00</u> | 0.95 | 0.04 | <u>0.00</u> |
| 6sr12p | 1.78 | 0.53 | <u>0.00</u> | 1.66 | 0.30 | <u>0.00</u> |
| All avg. | 1.07 | 1.09 | <u>0.15</u> | 0.94 | 0.96 | <u>0.05</u> |
| CC avg.    | 0.92 | 0.66 | <u>0.11</u> | 0.84 | 0.52 | <u>0.03</u> |
| CC p. avg. | 1.08 | 1.08 | <u>0.13</u> | 0.99 | 0.97 | <u>0.07</u> |
| CQ avg.    | 0.38 | 1.31 | <u>0.19</u> | 0.34 | 1.17 | <u>0.04</u> |
| CQ p. avg. | 0.34 | 0.84 | <u>0.06</u> | 0.33 | 0.80 | <u>0.04</u> |
| SR avg.    | 1.60 | 0.78 | <u>0.30</u> | 1.48 | 0.63 | <u>0.03</u> |
| SR p. avg. | 1.74 | 1.78 | <u>0.11</u> | 1.36 | 1.56 | <u>0.10</u> |
| 3-AP avg. | 0.34 | 1.15 | <u>0.10</u> | 0.30 | 0.95 | <u>0.03</u> |
| 4-AP avg. | 1.74 | 1.25 | <u>0.42</u> | 1.49 | 1.06 | <u>0.16</u> |
| 5-AP avg. | 1.22 | 1.48 | <u>0.07</u> | 1.07 | 1.39 | <u>0.00</u> |
| 6-AP avg. | 0.74 | 0.50 | <u>0.00</u> | 0.70 | 0.44 | <u>0.00</u> |